\documentclass[aps,prd,reprint,nofootinbib]{revtex4-1}

\usepackage[english]{babel}
\usepackage{hyphenat}
\usepackage{graphicx}
\usepackage{array}
\usepackage{bm}
\usepackage{amssymb}
\usepackage{amsmath}
\usepackage{amsfonts}
\usepackage{amsbsy}
\usepackage{latexsym}
\usepackage[colorlinks,citecolor=blue,urlcolor=blue,linkcolor=blue]{hyperref}
\usepackage{flushend}

\begin{document}

\title{Relativistic and nonrelativistic annihilation of dark matter: a sanity check \\
using an effective field theory approach
} 

\author{Mirco Cannoni }
\affiliation{Departamento de F\'isica Aplicada, Facultad de Ciencias Experimentales, Universidad de Huelva, 21071 Huelva, Spain}

\begin{abstract}
We find an exact formula for  the thermally averaged cross section times the relative velocity
$\langle \sigma v_{\text{rel}} \rangle$ with relativistic Maxwell-Boltzmann statistics.
The formula is valid in the effective field theory approach when the masses of the annihilation products 
can be neglected compared with the dark matter mass and cut-off scale.
The expansion at $x=m/T\gg 1$ directly gives the nonrelativistic
limit of $\langle \sigma v_{\text{rel}}\rangle$ which is usually used to compute the 
relic abundance for heavy particles that decouple when they are nonrelativistic. 
We compare this expansion with the one obtained by expanding the total cross section 
$\sigma(s)$ in powers of the nonrelativistic relative velocity $v_r$. We show the correct
invariant procedure that gives the nonrelativistic average $\langle \sigma_{nr} v_r \rangle_{nr}$
coinciding with the large $x$ expansion of $\langle \sigma v_{\text{rel}}\rangle$ in the comoving
frame.
We explicitly formulate flux, cross section, thermal average, collision integral of the 
Boltzmann equation in an invariant way using the true relativistic relative $v_\text{rel}$, 
showing the uselessness of the M\o{}ller velocity and further elucidating the conceptual and  numerical 
inconsistencies related with its use.
\end{abstract}
\maketitle

\section{Introduction}

While there are compelling evidences in astrophysics and cosmology that 
most of the mass of the Universe is composed by a new form of non baryonic
dark matter (DM), there is a lack of evidence for the existence of new physics at LHC 
and other particle physics experiments. 
On the theory side, many specific models with new particles and interactions beyond the standard 
model have been proposed to account for DM. 

Under these circumstances where no clear indications in favour of a particular model  
are at our disposal, the phenomenology of DM as been studied in a model independent 
way using an effective field theory approach, see for example 
\cite{Kurylov:2003ra,Fan:2010gt,Fitzpatrick:2012ix,Beltran:2008xg_vs2,Cao:2009uw_vs2,Buckley:2011kk_vs2,Matsumoto:2014rxa_vs2,Fedderke:2014wda_vs2_Moll,Chen:2013gya_vs2_Moll,Berlin:2014tja_vs2_error,Chang:2013oia_vs2,Cheung:2012gi_vs2,Balazs:2014rsa,Goodman:2010yf,Goodman:2010ku,Buchmueller:2013dya,Busoni:2013lha,Busoni:2014sya,Busoni:2014haa,Zheng:2010js_vl,Dreiner:2012xm,Blumenthal:2014cwa_vl,Desimone_moller_frame}.

Measurements of the parameters of standard model of cosmology~\cite{Bennett:2012zja,Ade:2013zuv}
furnish the present day mass density of DM, the relic abundance, $\Omega h^2 \sim 0.11$
with an uncertainty at the level of 1\%.
Any model that pretends to account for DM must reproduce this number, which, 
on the other hand, sets strong constraints on the free parameters of the model.
 
When the DM particles are  weakly interacting massive particles  that 
decouple from the primordial plasma at a temperature when they are nonrelativistic,
the relativistic averaged annihilation rate $\langle \sigma v_\text{rel}\rangle$ 
can be well approximated by taking the nonrelativistic average of the first two terms of the expansion of $\sigma$ in powers of the
nonrelativistic  relative velocity.  With $v_\text{rel}$ we indicate the {\em relativistic relative velocity} and with 
$v_r$ the {\em nonrelativistic relative velocity}, as defined in  \ref{app:2}. 
To describe collisions in a gas, and in particular in the primordial plasma,  
the reference frame that matters is the comoving frame (COF) where the observer 
sees the gas at rest as a whole and the colliding particles have general velocities
$\boldsymbol{v}_{1,2}$ without any further specification of the kinematics.

It is thus desirable to formulate cross sections and rates in a relativistic invariant
way, such that all the formulas and nonrelativistic expansions
are valid automatically in the COF. Obviously, invariant formulas give the same results in the 
lab frame (LF), the frame where one massive particle is at rest,
and in the center of mass frame (CMF) where the total momentum is zero. 
We will see that the key for the invariant formulation is $v_\text{rel}$.

On the contrary, in DM literature~\cite{GG} instead of $v_\text{rel}$ it is used
the so--called M\o{}ller velocity $\bar{v}$, see  \ref{app:2}. 
That this is incorrect was already discussed in Ref.~\cite{Cannoni:2013bza} 
but papers using $\bar{v}$ continue to appear. The problem with $\bar{v}$, which is not the 
relative velocity,
is its non invariant and nonphysical nature, for it can take values larger than $c$. 

In this paper we first find an exact formula for $\langle \sigma v_{\text{rel}} \rangle$
as a function of $x=m/T$ calculated with the relativistic Maxwell-Boltzmann 
statistics. The formula is valid in the effective field theory framework such that the masses 
of the annihilation products can be neglected compared with the DM and the cut-off scale.
For concreteness we work with fermion DM.
We find the thermal functions corresponding to various interactions
and in particular those corresponding to
$s$ and $p$ wave scattering in the nonrelativistic limit which is given by
the expansion at $x\gg 1$. 
This is done in Section~\ref{sec2}, and 
~\ref{app:1} contains
some mathematical results needed for the derivation of the exact formula and its asymptotic expansions.

Then, in Section \ref{sec3}, we present the correct invariant method for obtaining the same 
expansion by expanding the total annihilation cross section $\sigma(s)$ in powers of $v_r$.

We then discuss in Section~\ref{sec4}
the problems with the use $\bar{v}$, while the numerical impact on the relic abundance 
of some incorrect methods employed in literature is evaluated in Section~\ref{sec5}.

~\ref{app:2} is preparatory for the whole paper: we remind
how relativistic flux, cross section, rate, collision term of the Boltzmann equation and thermal averaged rate
can be defined in the invariant way in terms of $v_\text{rel}$ showing the uselessness of 
the M\o{}ller velocity.

\section{Exact formula for the thermal average in the effective approach}
\label{sec2}

\begin{figure}[t!]
\includegraphics*[scale=0.53]{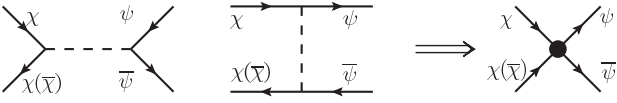}
\caption{$s$ and $t$ channel annihilation diagrams reducing to the effective vertex
corresponding to the lagrangian Eq.~(\ref{operator}). }
\label{Fig1}
\end{figure}

We consider a DM fermion field $\chi$ that couples to other fermion fields  
$\psi$ through an effective dimension-6 operator of the type
\begin{flalign}
\mathcal{L}_\Lambda=\frac{\lambda_a \lambda_b}{\Lambda^2} 
(\bar{\chi} \Gamma_a \chi) (\bar{\psi}\Gamma_b  \psi).
\label{operator}
\end{flalign}
The DM particles can be of Dirac or Majorana nature and
have mass $m$, while $\psi$
are the standard model fermions or new ones.
Here $\lambda_{a,b}$ are dimensionless  coupling associated with the interactions
described by combination of Dirac matrices $\Gamma_{a,b}$.
$\Lambda$ is the energy scale below which the effective field theory is valid.
In the exact theory $\Lambda$ corresponds to the mass of a heavy scalar or vector boson mediator that appears
in the propagators. 
The $\psi$ masses can be neglected compared to $\Lambda$ and $m$.
The exchange of a heavy 
mediator with mass $\Lambda$ may take place in  the $s$-channel 
and/or in $t$-channel, as depicted in Figure~\ref{Fig1},
depending on the specific model.

\subsection{Exact formula for $\langle \sigma {v}_{\text{rel}} \rangle$ }

In all generality, for $2 \to 2$ processes, the matrix elements depend only on 
two independent Mandelstam variables, for example $s$ and $t$, 
and the squared matrix element is dimensionless.
After integrating  over the CMF angle, for example, the only remaining 
dependence is on $s$ and $m$. Any amplitude related to the operator (\ref{operator}) 
gives an integrated squared matrix element
$\overline{|\mathcal{M}|^2}$ summed over the final spins  and
averaged over the initial spins
that is a simple polynomial of the type
\begin{flalign}
w=\int\overline{|\mathcal{M}|^2} d\cos\theta=
p_2 s^2 +p_1 m^2 s+p_0 m^4,
\end{flalign}   
with $p_0,...,p_2$ depending on $\Lambda$ and $\lambda_{a,b}$.
To get the formula for $\langle\sigma v_\text{rel}\rangle$ in a useful form,
it is convenient to define the {\it reduced cross section}
\begin{flalign}
\sigma_0 &=\frac{1}{2m^2}\frac{1}{32 \pi} w,
\label{sigma0}
\end{flalign}
and the {\it effective cross} section 
\begin{flalign}
\sigma_{{\Lambda}} =\frac{\lambda^2_a \lambda^2_b}{4\pi} \frac{m^2}{\Lambda^4},
\label{sigmalambda}
\end{flalign}
which contains all the couplings. 
In terms of the effective cross section (\ref{sigmalambda}), 
and of the dimensionless variable  $y=s/(4m^2)$,
the reduced cross section Eq.~(\ref{sigma0}) becomes
\begin{flalign}
\sigma_0=\sigma_\Lambda \left(a_2 y^2
+\frac{a_1}{4} y + \frac{a_0}{16}\right),
\label{sigma0ann}
\end{flalign}
where now $a_2,...,a_0$ are pure numbers.
The total unpolarized cross section then is   
\begin{flalign}
\sigma=\frac{2m^2}{ s} 
\frac{\sqrt{\lambda(s,m^2_3,m^2_4)}}{\sqrt{\lambda(s,m^2_1,m^2_2)}}
\sigma_0.
\label{sigma}
\end{flalign} 

We now set $m_1=m_2=m$, $m_3=m_4=0$ in Eq.~(\ref{sigma}) and in Eq.~(\ref{sigmav_1}),
and change variable to $y$. Thus Eq.~(\ref{sigmav_1}) becomes
\begin{flalign}
\langle \sigma {v}_{\text{rel}} \rangle
=\frac{2x}{K^2_2(x)}
\int_{1}^{\infty} {dy} \sqrt{y-1} K_1 (2x\sqrt{y}) \sigma_0(y).
\label{sigmav_y}
\end{flalign}
Using the integrals of ~\ref{app:1}, 
we find
\begin{flalign}
\langle \sigma {v}_{\text{rel}} \rangle=
\sigma_\Lambda \frac{1}{16}
[& 8 a_2 + 2 a_1 
 + (5a_2 +2a_1 +a_0)\frac{K^2_1(x)}{K^2_2(x)}
 \nonumber \\ 
 &+3a_2\frac{K^2_3(x)}{K^2_2(x)}].
\label{sigmav_final}
\end{flalign}
In the case $m_3=m_4=0$ the pure mass terms do not appear in 
the cross sections, thus $a_0 =0$. Furthermore, we can relate $a_2 $ and $a_1$ each other by an 
appropriate multiplicative factor, 
\begin{flalign}
a_1=k a_2,
\label{a_2_a_1}
\end{flalign}
and express the cross sections as a function of $a_2$ only.
The general formula (\ref{sigmav_final}) thus finally becomes
\begin{flalign}
\langle \sigma {v}_{\text{rel}} \rangle=
\sigma_\Lambda a_2 \mathcal{F}_k(x),
\label{sigmav_ann_simple}
\end{flalign}
with 
\begin{flalign}
\mathcal{F}_k(x) =\frac{1}{16}
\left(8 +2k + (5+2k)\frac{K^2_1(x)}{K^2_2(x)} + 3\frac{K^2_3(x)}{K^2_2(x)}\right)
\label{F_k}
\end{flalign}
the factored out thermal function.

The  nonrelativistic thermal average is given by the expansion at $x\gg 1$.
Using the asymptotic expansions Eq.~(\ref{Bessel}) we find
\begin{flalign}
\langle \sigma_{nr} v_r \rangle_{nr}=\sigma_\Lambda a_2 
\left(1+\frac{k}{4}
-\frac{3}{8} \frac{k}{x}\right)+\mathcal{O}(x^{-2}).
\label{nonrel}
\end{flalign} 

In the ultrarelativistic limit, $x\ll 1$, using the expansions (\ref{Bessel2}),
the thermal functions behave as 
$3/x^2$, thus 
\begin{flalign}
\langle \sigma {v}_{\text{rel}} \rangle_{ur} \sim
\sigma_\Lambda a_2 \frac{3}{x^2}=\frac{\lambda^2_a \lambda^2_b}{4\pi\Lambda^4}3a_2 T^2,
\end{flalign}
which is the expected result for massless particles.

The exact integration is possible because
the effective operator removes the momentum dependence in the propagators that are
reduced to a multiplicative constant and the assumption $m_3 =m_4 =0$ allows to simplify
the square root $\sqrt{\lambda(s,m^2_3,m^2_4)}=s$ in the cross section (\ref{sigma}).
For example, with $m_3=m_4=m_\psi$, equation (\ref{sigmav_y}) becomes
\begin{flalign}
\langle \sigma {v}_{\text{rel}} \rangle
=\frac{2x}{K^2_2(x)}
\int_{y_0}^{\infty} \hspace{-3mm}{{dy}\sqrt{y-\rho} \sqrt{y-1} K_1 (2x\sqrt{y}) \sigma_0(y,\rho)}.
\nonumber
\end{flalign} 
with $\rho=m^2_\psi / m^2$ and $y_0 =1$ if $m\geq m_\psi$, $y_0 =\rho$ if $m< m_\psi$. 
In this case the exact integration is not possible but nonrelativistic expansions 
exist also in the case $\rho=1$ and $\rho\gg 1$ as we have shown in Ref.~\cite{Cannoni:2013bza}.

\subsection{Applications}
 
In order to show the  thermal behaviour of different interactions, 
we calculate the cross sections 
for various operators of the type (\ref{operator}), both for 
 $s$ and $t$ channel annihilation.
We list the quantity $\varpi=\Lambda^4/(\lambda^2_a \lambda^2_b)\,w$
and the resulting average Eq.~(\ref{sigmav_ann_simple}).

For the \textit{$s$-channel annihilation} we find: \\
1) Scalar: $(\bar{\chi} \chi) (\bar{\psi} \psi)$, $(\bar{\chi} \chi) (\bar{\psi} \gamma^5 \psi)$.
\begin{flalign}
\varpi =2s(s-4m^2),\;\;\;
\langle \sigma_S {v}_{\text{rel}} \rangle  =\sigma_\Lambda 2 \mathcal{F}_{-4}(x).
\label{S_s}
\end{flalign}
2) Pseudo-scalar: $(\bar{\chi}\gamma^5 \chi) (\bar{\psi}\gamma^5  \psi)$, $(\bar{\chi}\gamma^5 \chi) (\bar{\psi} \psi)$:
\begin{flalign}
\varpi = 2s^2,\;\;\;
\langle \sigma_{PS} {v}_{\text{rel}} \rangle =\sigma_\Lambda 2 \mathcal{F}_{0}(x).
\label{PS_s}
\end{flalign}
3) Chiral: $(\bar{\chi}P_{L,R} \chi) (\bar{\psi} P_{L,R} \psi)$.
\begin{flalign}
\varpi =\frac{1}{2}s(s-2m^2),\;\;\;
 \langle \sigma_C {v}_{\text{rel}} \rangle =\sigma_\Lambda \frac{1}{2} 
\mathcal{F}_{-2}(x).
\label{C_s}
\end{flalign}
4) Pseudo-vector: $(\bar{\chi} \gamma^\mu \gamma_5 \chi) (\bar{\psi} \gamma_\mu \gamma_5\psi)$, 
$(\bar{\chi} \gamma^\mu \gamma_5 \chi) (\bar{\psi} \gamma_\mu \psi)$.
\begin{flalign}
\varpi&=\frac{8}{3}s(s-4m^2),\;\;\;
\langle \sigma_{PV} {v}_{\text{rel}} \rangle =\sigma_\Lambda \frac{8}{3} \mathcal{F}_{-4}(x)
\label{PV_s}
\end{flalign}
5) Vector: $(\bar{\chi} \gamma^\mu \chi) (\bar{\psi} \gamma_\mu \psi)$, $(\bar{\chi} \gamma^\mu \chi)(\bar{\psi} \gamma_\mu \gamma^5 \psi)$.
\begin{flalign}
\varpi=\frac{8}{3}s(s+2m^2),\;\;\;
\langle \sigma_{V} {v}_{\text{rel}} \rangle =\sigma_\Lambda \frac{8}{3} \mathcal{F}_{2}(x).
\label{V_s}
\end{flalign}
6) Vector-chiral: $(\bar{\chi} \gamma^\mu P_{L,R} \chi) (\bar{\psi} \gamma_\mu P_{L,R} \psi)$.
\begin{flalign}
\varpi =\frac{8}{3}s(s-m^2),\;\;\;
\langle \sigma_{VC} {v}_{\text{rel}} \rangle  =\sigma_\Lambda \frac{8}{3} \mathcal{F}_{-1}(x).
\label{VC_s}
\end{flalign}
The tensor interaction $\sigma^{\mu \nu}$ gives the same function as the vector case and is not
reported. In the case of a Majorana $\chi$ clearly the vector and tensor interactions are absent,
and the inclusion of a factor $1/2$ in the operator (\ref{operator}) cancels the factor 4 due to 
the presence of the exchange diagram of the initial identical particles.

Now we consider some examples of \textit{$t$-channel annihilation} 
for operators common to Dirac and  Majorana DM annihilation:\\
1) Scalar, pseudo-scalar:
$(\bar{\chi} \chi) (\bar{\psi} \psi)$, $(\bar{\chi} \chi) (\bar{\psi} \gamma^5 \psi)$,\\
$(\bar{\chi}\gamma^5 \chi) (\bar{\psi}\gamma^5  \psi)$, $(\bar{\chi}\gamma^5 \chi) (\bar{\psi} \psi)$.
\begin{flalign}
\varpi_D&=\frac{2}{3}s(s-m^2), \;\;\; 
\langle \sigma^{D,t}_{S,PS}\, {v}_{\text{rel}} \rangle  =\sigma_\Lambda  \frac{2}{3} \mathcal{F}_{-1}(x). \\
\varpi_M& =\frac{1}{3}s(5s-14m^2), \;\;\; 
\langle \sigma^{M,t}_{S,PS}\, {v}_{\text{rel}} \rangle  =\sigma_\Lambda  \frac{1}{3} \mathcal{F}_{-\frac{14}{5}}(x). 
\end{flalign}
2) Chiral: $(\bar{\chi}P_{L,R} \chi) (\bar{\psi} P_{L,R} \psi)$.
\begin{flalign}
\varpi_D& =\frac{1}{6}s(s-m^2), \;\;
\langle \sigma^{D,t}_{C}\, {v}_{\text{rel}} \rangle =\sigma_\Lambda  \frac{1}{6}
\mathcal{F}_{-1}(x). \\
\varpi_M &=\frac{1}{3}s(s-4m^2), \;\;
\langle \sigma^{M,t}_{C}\, {v}_{\text{rel}} \rangle =\sigma_\Lambda  \frac{1}{3}
\mathcal{F}_{-4}(x).
\end{flalign}
3) Pseudo-vector:
$(\bar{\chi} \gamma^\mu \gamma_5 \chi) (\bar{\psi} \gamma_\mu \gamma_5\psi)$, 
$(\bar{\chi} \gamma^\mu \gamma_5 \chi) (\bar{\psi} \gamma_\mu \psi)$.
\begin{flalign}
\varpi_D &= \frac{4}{3}s(4s-7m^2), \;\;
\langle \sigma^{D,t}_{PV} {v}_{\text{rel}} \rangle =\sigma_\Lambda  \frac{4}{3} \mathcal{F}_{-\frac{7}{4}}(x). \\
\varpi_M &=\frac{8}{3}s(7s-16m^2), \;\;
\langle \sigma^{M,t}_{PV} {v}_{\text{rel}} \rangle =\sigma_\Lambda  \frac{8}{3} \mathcal{F}_{-\frac{16}{7}}(x). 
\end{flalign}
\begin{figure}[t!]
\includegraphics*[scale=0.38]{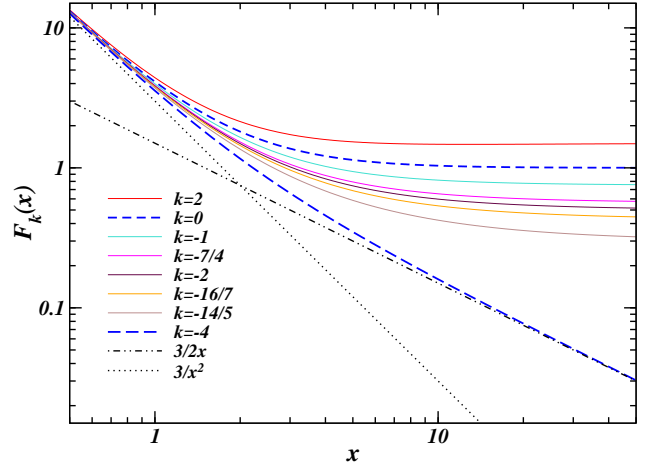}
\caption{The thermal function (\ref{F_k}) for the interactions and annihilation cross sections
considered in the text.}
\label{Fig2}
\end{figure}

The thermal functions corresponding to the previous cases are shown in Figure~\ref{Fig2}
where the asymptotic behaviours are clearly seen.
In particular we note that 
\begin{flalign}
\mathcal{F}_0 (x)&=
\frac{1}{16}
\left(8 + 5\frac{K^2_1(x)}{K^2_2(x)} + 3\frac{K^2_3(x)}{K^2_2(x)}\right),\\
\mathcal{F}_{-4}(x)&= 
\frac{3}{16}
\left(-\frac{K^2_1(x)}{K^2_2(x)} + \frac{K^2_3(x)}{K^2_2(x)}\right),
\end{flalign}
behave in the nonrelativistic limit as
\begin{flalign}
\mathcal{F}_0 (x)
\sim 1 +\mathcal{O}(x^{-2})\nonumber,\;\;\;
\mathcal{F}_{-4}(x) 
\sim \frac{3}{2x} +\mathcal{O}(x^{-2})\nonumber.
\end{flalign}
The function $\mathcal{F}_0(x)$, which appears in the $s$-channel
annihilation through a pseudoscalar interaction, is the only case where 
the term of order $\mathcal{O}(x^{-1})$  is absent, while    
$\mathcal{F}_{-4}(x)$, which appears in the scalar and axial-vector $s$-channel annihilation and 
in the chiral $t$-channel Majorana fermion annihilation, is the only case where the constant $\mathcal{O}(x^{0})$ term is zero.
These are the exact temperature dependent factors
that correspond to the phenomenological interpolating functions 
proposed in Ref.~\cite{Drees:2009bi} to model the  $s$-wave and $p$-wave behaviour in the 
nonrelativistic limit. For all other interactions both $s$-wave and $p$-wave contribution 
are present. The function $\mathcal{F}_{-4}(x)$ can be also read off from the formulas 
of Ref.~\cite{Claudson:1983js} where the $t$-channel annihilation of Majorana fermions  
with the exchange of a scalar with chiral couplings was considered.

We note that although we have concentrated on the case of fermion DM, 
the formula is valid for DM scalar and vector candidates as well, with the necessary redefinition of 
$\sigma_\Lambda$.

\section{Expansion of the cross section in powers of the relative velocity}
\label{sec3}

In the general case $m_3=m_4=m_\psi\neq 0$ the exact integration is not possible. If the
relative velocity of the annihilating particles is small compared with the velocity of light 
we can work directly with nonrelativistic formulas.
The exothermic annihilation cross section in the nonrelativistic limit, to the lowest
orders in $v_r$, is usually expanded as $\sigma_{nr} \sim a/v_r +b v_r$, and multiplying by $v_r$,
\begin{flalign}
\sigma_{nr} v_r \sim a+b v^2_r.
\label{sigmav_a_b}
\end{flalign}
Then,  using Eq.~(\ref{Maxwell_vrel}) and (\ref{nr_average_def}), the nonrelativistic
thermal average of Eq.~(\ref{sigmav_a_b}) is 
\begin{flalign}
\langle \sigma_{nr} v_r \rangle_{nr}  
\sim a + 6 \frac{b}{x}.
\label{sigmav_avea_a_b}
\end{flalign}

In the case of our cross sections, comparing Eq.~(\ref{sigmav_avea_a_b}) 
with Eq.~(\ref{nonrel}), the coefficients are thus~\footnote{This result must coincide with the 
expansion of Ref.~\cite{SWO,Cannoni:2013bza}.  
With our notation the expansion is 
\begin{flalign}
\langle \sigma_{nr} v_r \rangle_{nr}\sim
\sigma_0|_{y=1}
+\frac{3}{x}\left(-\sigma_0|_{y=1}+\frac{1}{2}\sigma'_0|_{y=1}\right),
\nonumber
\end{flalign}
where the prime indicate derivative respect to the variable $y$.
Comparison with the expansion 
(\ref{sigmav_avea_a_b}) requires to identify
\begin{flalign}
a\equiv\sigma_0|_{y=1}, \;\;\;
b\equiv\frac{1}{2}\left(\frac{1}{2}{\sigma_0 '|_{y=1}} -\sigma_0|_{y=1}\right).
\nonumber
\end{flalign}
Using Eq.~(\ref{sigma0ann}) with $a_0 =0$ and  $a_2 =k a_1$, it is easy 
to verify that one obtains again Eq.~(\ref{nonrel}).}
\begin{flalign}
a =\sigma_\Lambda a_2 \left(1+\frac{k}{4}\right),\;\;\;
b =-\sigma_\Lambda a_2 \frac{k}{16}.
\label{a_b_corr_gen}
\end{flalign}

We now ask, given $\sigma(s)$, how to perform the expansion in terms of the relative 
velocity to find the coefficients $a$ and $b$ that correspond to the large $x$ expansion 
of the relativistic thermal average in the COF.
Combining equations (\ref{sigma0ann}), (\ref{sigma}), (\ref{a_2_a_1}), the general 
total annihilation cross section reads
\begin{flalign}
\sigma=\sigma_\Lambda \frac{a_2}{2} \frac{\sqrt{s}}{\sqrt{s-4m^2}} \left(\frac{s}{4m^2} +\frac{k}{4}\right).
\label{sigmatot_k}
\end{flalign}
The correct way to proceed is to use the invariant relation Eq.~(\ref{vrel_s})
with $m_1= m_2=m$ and to solve it for $s$ as a function of $v_{\text{rel}}$:
\begin{flalign}
s=2m^2 \left(1+\frac{1}{\sqrt{1-v^2_\text{rel}}}\right).
\label{s_vrel}
\end{flalign}
This formula is valid in every frame and substituted in  Eq.~(\ref{sigmatot_k}) gives the exact dependence 
of the cross section on the relativistic relative velocity, $\sigma(v_\text{rel})$. 
Then, if $v_\text{rel}\sim v_r \ll 1$,
we can expand the obtained expression to the desired order in $v_r$ and the nonrelativistic 
average taken using Eq.~(\ref{nr_average_def}) will coincide with the expansion of Eq.~(\ref{sigmav_ann_simple}) for $x\gg 1$,
that is the expansion (\ref{nonrel}).

Equivalently, in order to find the expansion (\ref{sigmav_a_b}),  we note that 
the squared roots in the annihilation cross section 
(\ref{sigmatot_k})
imply that a term of order $v^4_r$ in $s$ will contribute to the order $v^2_r$ in $\sigma$. 
Thus we need to expand $s$, formula (\ref{s_vrel}), at least to order $v^4_r$, 
\begin{flalign}
s \sim 4m^2+m^2 v^2_r +\frac{3}{4} m^2 v^4_r.
\label{s_expansion_vrel}
\end{flalign}
Substituting Eq.~(\ref{s_expansion_vrel}) in Eq.~(\ref{sigmatot_k}) 
and performing the expansion  in powers of $v_r$ it easy
to find
\begin{flalign}
{\sigma}_{nr} v_{r} \sim \sigma_\Lambda {a_2} \left(1+\frac{k}{4}-\frac{k}{16}v^2_{r}
\right),
\label{expansion__sigmatot_k}
\end{flalign}
in  agreement with (\ref{a_b_corr_gen}).

In the case of coannihilations~\cite{Griest:1990kh}, for example when a DM particles
scatter off another particle with different mass, the Mandelstam invariant
takes the form
\begin{flalign}
s=(m_1 -m_2)^2+ 2m_1 m_2 \left(1+\frac{1}{\sqrt{1-v^2_\text{rel}}}\right),
\label{s_vrel_coan}
\end{flalign}
with the expansion
\begin{flalign}
s \sim (m_1-m_2)^2+m_1 m_2 v^2_r +\frac{3}{4} m_1 m_2 v^4_r.
\label{s_expansion_vrel_coan}
\end{flalign}
This procedure gives the correct 
expansion in the  COF where the velocities 
$\boldsymbol{v}_{1,2}$ of the colliding particles are specified in this frame.
Clearly, the same expansion with the same coefficients is obtained in the LF and in the CMF.

\section{The problems with the M\o{}ller velocity}
\label{sec4}

The simple outlined  procedure has not been recognized in DM literature
where, incorrectly,  
the M\o{}ller velocity $\bar{v}$, 
Eq.~(\ref{v_moller}), instead of $v_\text{rel}$ is considered.
As reminded in ~\ref{app:2}, $\bar{v}$ is a non-invariant,
non-physical velocity. 
The expression of $\bar{v}$ in terms of $s$ is thus different in different frames
and the expansion of
$\sigma$ takes different values in different frames.

Before discussing the problems with the M\o{}ller velocity we note that if we take the 
limit $m_f\to 0$ in the analogous expansions published many papers 
~\cite{Beltran:2008xg_vs2,Cao:2009uw_vs2,Buckley:2011kk_vs2,Matsumoto:2014rxa_vs2,Fedderke:2014wda_vs2_Moll,Chen:2013gya_vs2_Moll,Berlin:2014tja_vs2_error,Chang:2013oia_vs2},
we do not reproduce the expansion (\ref{expansion__sigmatot_k}). 
The reason is that in these papers the expansion of $s$ is truncated to
the lowest order in $v^2_{r}$,
\begin{flalign}
s \sim 4m^2 + m^2 v^2_{r}.
\end{flalign}
If we substitute this in Eq.~(\ref{sigmatot_k}) and expand, we find
\begin{flalign}
\sigma_{nr} v_{r} \sim \sigma_\Lambda 
{a_2}\left(1+\frac{k}{4}+\frac{12+k}{32}v^2_{r}
\right),
\label{sv_incorretta1}
\end{flalign}
with an incorrect coefficient $b$.  
Clearly the same wrong result is obtained truncating (\ref{s_expansion_vrel}) to order
$v^2_r$, whatever the frame in which $v_r$ is specified, CMF, LF or COF.

We now go back to the M\o{}ller velocity (\ref{v_moller}).
Evaluated in the CMF  
taking $m_1 =m_2 =m$ reads
\begin{flalign}
\bar{v}_* =\frac{2}{\sqrt{s_*}} \sqrt{s_* -4 m^2} .
\label{v_moller_cm}
\end{flalign}
We indicate the quantities evaluated in the CMF with  a "*".
By inverting Eq.~(\ref{v_moller_cm}) we find
\begin{flalign}
s_* =\frac{4 m^2}{1-\frac{\bar{v}^2_{*}}{4}}.
\label{s_vr_cmf}
\end{flalign}
This relation is different from (\ref{s_vrel})
and is often incorrectly identified 
as the relation between $s$ and the relative velocity in the CMF, see for 
example \cite{Griest:1990kh},~\cite{Berlin:2014tja_vs2_error}. 
In facts, the expansion to order $\mathcal{O}(v^4_{r,*})$ reads
\begin{flalign}
s_* \sim 4m^2 + m^2 v^2_{r,*}+\frac{m^2}{4} v^4_{r,*}.
\label{s_vr_cmf_1}
\end{flalign}
When used in (\ref{sigmatot_k}), it gives the following nonreltivistic expansion
of the cross section 
\begin{flalign}
\sigma_{nr} v_{r,*} \sim \sigma_\Lambda 
{a_2}\left(1+\frac{k}{4}+\frac{1}{4}v^2_{r,*}
\right),
\label{sv_incorretta2}
\end{flalign}
which is different from the correct expansion (\ref{expansion__sigmatot_k}).

Other authors,~\cite{GG} and \cite{Zheng:2010js_vl,Dreiner:2012xm,Blumenthal:2014cwa_vl,Desimone_moller_frame}, 
perform the expansion with the M\o{}ller
velocity evaluated in the rest frame of one particle. Indicating with "$\ell$" the quantities in this frame,
Eq.~(\ref{v_moller}) becomes
\begin{flalign}
\bar{v}_\ell =\frac{\sqrt{s_\ell} \sqrt{s_\ell -4 m^2}}{s_\ell -2m^2},
\label{v_moller_lab}
\end{flalign}
and by inverting Eq.~(\ref{v_moller_lab}) we obtain
\begin{flalign}
s_\ell=2m^2 \left(1+\frac{1}{\sqrt{1-\bar{v}^2_\ell}}\right).
\end{flalign}
This expression is formally identical to Eq.~(\ref{s_vrel}), thus 
when $\bar{v}_\ell \sim v_{r,\ell}$ and $s_\ell$ is expanded  up to the order $v^4_{r,\ell}$ 
we obtain the expansion $\sigma_{nr} v_{r,\ell}$ which formally coincides
with Eq.~(\ref{expansion__sigmatot_k}), with $v_{r,\ell}$ in place of $v_r$.

It should be clear that this is just
a mathematical coincidence due to the fact that $\bar{v}$ 
reduces to $v_\text{rel}$ only when one of the two velocities 
$\boldsymbol{v}_{1,2}$ is zero as it is evident from the definitions 
Eq.~(\ref{v_rel_Rel_def}) and Eq.~(\ref{v_moller}).
In other words, the expansion found in Refs.~\cite{Zheng:2010js_vl,Dreiner:2012xm,Blumenthal:2014cwa_vl,Desimone_moller_frame}
are correct because the authors have implicitly used the relative velocity, Eq.~(\ref{vrel_s}) and (\ref{s_expansion_vrel}).

We thus  emphasize some common statements
found in DM literature and why they do not subsist:\\ 
1) \textit{ In the relativistic Boltzmann equation the $v$ in
$\sigma v$ is $\bar{v}$ and $\langle \sigma v\rangle$ must be calculated in the LF frame.}\\
This is not true, as shown in details in Ref.~\cite{Cannoni:2013bza} and in ~\ref{app:2}.
Using $v_\text{rel}$ and recognizing the nonphysical nature of $\bar{v}$,
one works always with invariant quantities and the consistency of the  relativistic and nonrelativistic 
formulas and expansions is obtained in the comoving frame
without any further specification of the kinematics.
The LF, also called M\o{}ller frame in Ref.~\cite{Desimone_moller_frame}, cannot be a  
privileged frame for the relic abundances calculation also because for massless particles the rest frame does not exist.\\ 
2) \textit{ The M\o{}ller velocity coincides 
with relative velocity in a frame where the velocities are collinear.}\\
This not true because, for example,
in the CMF where the particles have velocities $v_*$, the M\o{}ller velocity is $2v_*$ while the relative velocity is 
$2v_* /(1+v^2_*)$. Note that the true relative velocity is never superluminal.

\section{Impact on the relic abundance }
\label{sec5}

Only in the case $k=-4$ the incorrect expansions (\ref{sv_incorretta1}) and (\ref{sv_incorretta2})
coincide, incidentally, with the expansion (\ref{expansion__sigmatot_k}).
While the lowest order coefficient $a$ turns out to be always
the same, the coefficient $b$ is different in any other case.
To illustrate the impact of $b$ on the value of the relic abundance
we consider the case of the $s$-channel annihilation 
with vector interaction, Eq.~(\ref{V_s}), and the $s$-channel annihilation
with a pseudoscalar exchange, Eq.~(\ref{PS_s}). 
In the first case $k=2$, $a_2 =8/3$, and
the correct coefficients $a$ and $b$ are
\begin{flalign}
a_V= 4 \sigma_\Lambda, \;\;\;
b_V=- \frac{\sigma_\Lambda}{3},
\label{a_b_V_corr}
\end{flalign}
while the incorrect coefficient $b$ in 
(\ref{sv_incorretta1}) and (\ref{sv_incorretta2}) is
\begin{flalign}
b_{V_1}= \frac{7}{6} \sigma_\Lambda,\;\;\;\;
b_{V_2}= \frac{2}{3} \sigma_\Lambda.
\label{b_V_wrong}
\end{flalign}
In the second case, $k=0$ and  $a_2 =2$, thus
\begin{flalign}
a_{PS}= 2 \sigma_\Lambda, \;\;\;
b_{PS}=0,
\label{a_b_PS_corr}
\end{flalign}
and the wrong $b$ coefficients are
\begin{flalign}
b_{PS_1}= \frac{3}{4} \sigma_\Lambda,\;\;\;\;
b_{PS_2}= \frac{1}{2} \sigma_\Lambda.
\label{b_PS_wrong}
\end{flalign}

We calculate the relic abundance following the exact theory of freeze out presented
in Ref.~\cite{Cannoni:2014zqa}.  We briefly recall the main points.
Let $Y_{0}=45/(4 \pi^4)(g_\chi /g_s)
x^2 K_2 (x)$ be the initial equilibrium abundance (number density over the entropy density),
with $g_\chi =2$ for spin 1/2 fermions and $g_s$ the relativistic degrees of freedom associated
with the entropy density.
The function $Y_1(x)$ that gives the abundance up to the point $x_*$ where $Y_1(x)-Y_0 (x)$
is maximal is
\begin{flalign}
Y_1(x)&=(1+\delta(x))Y_0 (x),
\label{Y1}
\\
\delta (x)&=\sqrt{1- 
\frac{x^2}{C \langle \sigma v_\text{rel}\rangle Y_{0}  }
\frac{1}{Y_0} \frac{dY_{0}}{dx}} -1,
\label{delta}
\end{flalign}
with $x_\ast$ given by the condition
\begin{flalign}
-\frac{1}{Y_0(x)}\frac{dY_0(x)}{d x}=
\frac{1}{\delta(x)} \frac{d\delta(x)}{dx}\,\, \text{at $x=x_\ast$}.
\label{freezeoutcondition}
\end{flalign}
The abundance at $x>x_\ast$ is found by integrating numerically the usual equation 
\begin{flalign}
\frac{dY}{dx}= 
\frac{C}{x^2} {\langle \sigma v_\text{rel}\rangle}(Y^2_{0}-Y^2), 
\label{Beq}
\end{flalign}
with the initial condition ($x_*$, $Y(x_*)=Y_1 (x_*)$).
The factor $C$ is defined by
$C =\sqrt{\frac{\pi}{45}} M_P m_\chi \sqrt{g_*}$,
where $M_P$ is the Plank mass and 
$\sqrt{g_*} ={g_s}/ {\sqrt{g_\rho}} (1+{T}/{3}\,  d(\ln g_s )/dT)$
accounts for the temperature dependence of the relativistic degrees of freedom 
associated with the energy density, $g_\rho$,  and 	 $g_s$~\cite{SWO,GG}. 
For WIMP masses larger than 10 GeV we can neglect 
the temperature dependence of the degrees of freedom~\cite{Steigman:2012nb,Drees:2015exa} and
take $g_s =g_\rho=g=100$, $\sqrt{g_*}=\sqrt{g}$.
In solving numerically (\ref{Beq}) and (\ref{freezeoutcondition}) with the exposed method, 
we use the exact formula for  
$\langle \sigma v_\text{rel}\rangle$, Eq.~(\ref{sigmav_ann_simple}).

We compare the previous numerical solution with the one obtained using
the nonrelativistic freeze out approximation (FOA) that is commonly employed
in literature. The FOA consists in integrating equation (\ref{Beq}) 
with an initial condition ($x_f$, $Y(x_f)$)
such that the equilibrium term proportional $Y^2_{0}$ can be neglected.
We choose the freeze out point
at the point $x_2$ where $Y(x_2)\simeq Y_1(x_2)=2Y_0 (x_2)$. As shown in  
Ref.~\cite{Cannoni:2014zqa},
$Y_1 (x)$ well approximates the true abundance also in the interval  $x_* <  x < x_2 $.
$x_2$ is the optimal point for the FOA and
corresponds to the temperature where the extent of the inverse creation reaction $\psi\bar{\psi}\to \chi\chi$ is maximal.
The solution in the freeze out approximation is then
\begin{flalign}
Y_{FOA}=\frac{ 2 Y_0 (x_2)}{1+ 2 Y_0(x_2) \frac{C}{x_2}(a+3\frac{b}{x_2})}.
\label{Y2inf}
\end{flalign}
The freeze out point $x_2$ is given
by the condition $-\frac{1}{Y_0} \frac{dY_0}{dx}\\=3 \frac{C}{x^2} \langle \sigma v_\text{rel}  \rangle Y_0$,
which, in terms of the  method of Ref.~\cite{Scherrer:1985zt} 
corresponds to $c(c+2)=3$, that is $c=1$. 
Using the nonrelativistic form of $Y_0$,
\begin{flalign}
Y_{0}= \frac{45}{4 \pi^4}\frac{g_\chi}{g_s}\sqrt{\frac{\pi}{2}}
x^{3/2} e^{-x},
\label{Y_0}
\end{flalign}
$x_2$ is given by the root of
\begin{flalign}
3 {C} \left(a+6 \frac{b}{x} \right)\sqrt{\frac{\pi}{2}}x^{-1/2}e^{-x}=1.
\label{x2mio_general}
\end{flalign}
Calling $\alpha=3a C \sqrt{\pi/2} $, an accurate analytical approximate solution of Eq.~(\ref{x2mio_general}) is given by
\begin{flalign}
x_2 = \ln\alpha
-\frac{1}{2}\ln(\ln\alpha)
+\ln(1+\frac{6 b}{a}(\ln\alpha)^{-1}).
\label{x2}
\end{flalign}
\begin{figure}[t!] 
\includegraphics*[scale=0.43]{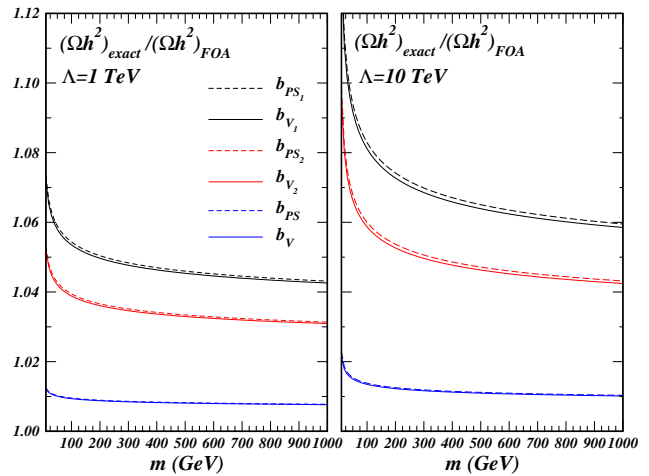}
\caption{Ratio of the relic abundance obtained by solving numerically
equation (\ref{Beq}) over the value given by the freeze out
approximation, for the pseudoscalar and vector interactions. 
In the bottom blue curves for the FOA the correct coefficients (\ref{a_b_V_corr}) and (\ref{a_b_PS_corr})
are used. The red and the black curves show the effect of the wrong coefficients (\ref{b_V_wrong}) and (\ref{b_PS_wrong}),
respectively. 
}
\label{Fig3}
\end{figure}

The relic abundance normalized over the critical density is 
$\Omega h^2 = 2.755\times 10^8 (m/\text{GeV}) Y_{(\infty)}$  
for a Majorana fermion and two times that quantity for a Dirac fermion with the same density of antiparticles.
We now compare the exact relic abundance $\Omega h^2$ with the value $(\Omega h^2)_{FOA}$
furnished by the nonrelativistic FOA calculated using the correct and the wrong expansions. 
We take the couplings $\lambda_{a,b}=1$ for illustrative purposes
and two values of the cut off scale, $\Lambda=1,10$ TeV.
The value of the freeze out points $x_\ast$ and $x_2$ varies roughly
between 18 and 30 in the parameter space with $m<\Lambda$ where the effective 
treatment is supposed to be valid. 
The ratio $\Omega h^2 / (\Omega h^2)_{FOA}$ is shown in Figure \ref{Fig3} as a function of the 
DM mass for the chosen examples.
The bottom blue curves  show that the FOA with the correct coefficients (\ref{a_b_V_corr}) and (\ref{a_b_PS_corr})
underestimates the numerical value by less than 2\%, and that in  most part of the parameter space the error 
is at the level of 1\% or less. This a test of goodness for our FOA, and confirms what shown in Ref.~\cite{Cannoni:2014zqa}.
 The red and the black curves show the effect of the wrong coefficients (\ref{b_V_wrong}) and 
(\ref{b_PS_wrong}), respectively.
The wrong expansions underestimate the relic abundance 
by a factor between $3\%$ and $12\%$ for both interactions for masses larger than 10 GeV
as shown in the plot. The behaviour is similar for the other interactions not
shown in figure. The error becomes even larger at smaller masses and
we have verified that using for example $c=1/2$ and other values we get even worst approximations.
Clearly this kind 
of error nowadays is not compatible with the precision with which the experimental value
is known. 
%
%

\begin{acknowledgements}
This work was supported by a grant under the MINECO/FEDER project: SOM: Sabor y origen de la Materia (CPI-14-397).
The author acknowledges Roberto Ruiz de Austri, Nuria Rius and Pilar Hernandez for hospitality and 
for useful discussions at the Instituto de Fisica Corpuscolar (IFIC) in Valencia where part  of this 
work was done.  
\end{acknowledgements}

\appendix

\section{Integrals and expansions }
\label{app:1}

Equation (\ref{sigmav_y}) can be written as 
\begin{flalign}
\langle \sigma {v}_{\text{rel}} \rangle
=\sigma_\Lambda 
\frac{2x}{K^2_2(x)}
\left(a_2 \mathsf{A}_2 + \frac{a_1}{4} \mathsf{A}_1 +\frac{a_0}{16} \mathsf{A}_0\right).
\label{Ap2_1}
\end{flalign}
The integrals are  evaluated with methods similar to those described in Ref.~\cite{Cannoni:2013bza}
in terms of Bessel functions of the second kind: 
\begin{flalign}
\mathsf{A}_0&=\int_{1}^{\infty} {dy} \sqrt{y-1} K_1 (2x\sqrt{y}) 
=\frac{1}{2x} K^2_1(x),
\nonumber
\\
\mathsf{A}_1 &=\int_{1}^{\infty} {dy} \sqrt{y-1}y K_1 (2x\sqrt{y})  
= \frac{1}{2x} \frac{K^2_1(x)+ K^2_2(x)}{2},
\nonumber
\\
\mathsf{A}_2 &=\int_{1}^{\infty} {dy} \sqrt{y-1}y^2 K_1 (2x\sqrt{y}) \crcr
&= \frac{1}{2x}\frac{1}{16} [5 K^2_1(x)+ 8 K^2_2(x) +3K^2_3(x)].
\nonumber
\end{flalign}
The expansions at $x\gg 1$ are
\begin{flalign}
\frac{K^2_1(x)}{K^2_2(x)} \sim 
1-\frac{3}{x}+\mathcal{O}(x^{-2}),\;\; 
\frac{K^2_3(x)}{K^2_2(x)} \sim 
1+\frac{5}{x}+\mathcal{O}(x^{-2}), 
\label{Bessel}
\end{flalign}
while for $x\ll 1$ are
\begin{flalign}
\frac{K^2_1(x)}{K^2_2(x)} \sim 
\frac{x^2}{4}+\mathcal{O}(x^{3}), \;\;\;
\frac{K^2_3(x)}{K^2_2(x)} \sim 
\frac{16}{x^2} +\mathcal{O}(x^{2}).
\label{Bessel2}
\end{flalign}

\section{Invariant formulation using $v_\textbf{rel}$}
\label{app:2}

In this Appendix we remind, based on the results of Ref.~\cite{Cannoni:2013bza}, 
the main points about the relation between the relative velocity, the M\o{}ller velocity, flux and
thermal average which are used in the main text. 

\subsection{Invariant relative velocity }

The relativistic relative velocity that
generalizes the nonrelativistic relative velocity 
\begin{flalign}
v_r=|\boldsymbol{v}_{1}-\boldsymbol{v}_{2}|,
\label{v_r_def}
\end{flalign} 
is given by 
\begin{flalign}
v_{\text{rel}}=
\frac{
\sqrt{(\boldsymbol{v}_1 - \boldsymbol{v}_2)^2 - 
\frac{(\boldsymbol{v}_1 \times \boldsymbol{v}_2)^2}{c^2}
}}
{1-\frac{\boldsymbol{v}_1 \cdot \boldsymbol{v}_2}{c^2}}.
\label{v_rel_Rel_def}
\end{flalign} 
We have explicitly written the dependence on the velocity of light $c$ to make manifest
that $v_{\text{rel}}$ coincide with $v_r$ in the nonrelativistic limit
because the scalar and vector products are of order $(v/c)^2$. 
In the following we go back to natural units.

The relative velocity $v_{\text{rel}}$  
can be written using the Mandelstam invariant  $s=(p_1+p_2)^2$, where $p_{1,2}$ are the four-momenta,
and $\lambda$, the Mandelstam triangular function,
\begin{flalign}
\lambda(s,m^2_1,m^2_2)=[s-(m_1 + m_2)^2][s-(m_1 - m_2)^2],
\label{Mandelstam_l}
\end{flalign}   
in a generic frame,
\begin{flalign}
v_{\text{rel}}&=\frac{\sqrt{(p_1 \cdot p_2)^2 -m^2_1 m^2_2}}{p_1 \cdot p_2}
\label{vrel_mom_rep}\\
&=\frac{\sqrt{\lambda(s,m^2_1,m^2_2)}}    {s-(m^2_1+m^2_2)},
\label{vrel_s}
\end{flalign}
showing its invariant nature.

\subsection{Flux factor}
Given two bunches of particles with number densities $n_{1,2}$
and velocities $\boldsymbol{v}_{1,2}$ in a generic inertial frame, 
in nonrelativistic physics the flux is $F_{nr}=n_1 n_2 v_r$. To obtain
the relativistic invariant flux that reduces to $F_{nr}$ in the nonrelativistic limit,
the easiest way is to consider the 4-currents 
$J_i=(n_i, n_i \boldsymbol{v}_i)$, thus
\begin{equation}
F=(J_1 \cdot J_2) v_\text{rel}=n_1 n_2 (1-\boldsymbol{v}_1 \cdot \boldsymbol{v}_2)  v_{\text{rel}}.
\label{invF} 
\end{equation}
Note that the factor $(1-\boldsymbol{v}_1 \cdot \boldsymbol{v}_2)$ that guarantees
the Lorentz invariance of the product of the number densities can also be written 
as
\begin{flalign}
1-\boldsymbol{v}_1 \cdot \boldsymbol{v}_2=\frac{\gamma_{\text{r}}}{\gamma_1 \gamma_2}=\frac{p_1\cdot p_2}{E_1 E_2},
\label{fac}
\end{flalign}
where $\gamma_{{\text{r}}}=1/\sqrt{1-v^2_{\text{rel}}}$ is the Lorentz factor associated
with $v_\text{rel}$ and $\gamma_i$ the Lorentz factors associated with $\boldsymbol{v}_i$.

If the element of Lorentz invariant phase space is defined as usual
\begin{flalign}
d\tilde{p_i}=\frac{d^3 \boldsymbol{p}_i}{(2\pi)^3 2E_i},
\label{dlips}
\end{flalign}
and one particle states for bosons and fermions are normalized to $2E_i$ 
such that the density per unit volume is $2E_i$,
then, using (\ref{fac}), the flux (\ref{invF}) simplifies to
\begin{equation}
F= 4 (p_1\cdot p_2) v_{\text{rel}}.
\label{invF1} 
\end{equation}
Substituting the expression of $v_{\text{rel}}$ in the momentum representation, 
formula (\ref{vrel_mom_rep}), in Eq.~(\ref{invF1}), the scalar product $p_1 \cdot p_2$
cancels out and the standard explicit form is recovered
\begin{equation}
F=4{\sqrt{(p_1 \cdot p_2)^2 -m^2_1 m^2_2}}.
\end{equation}

\subsection{Cross section and collision integral}

The integrated collision term of the Boltzmann equation, neglecting quantum effects, can be written as,  
\begin{flalign}
\int \prod^4_{i=1} d\tilde{p_i}\;
[f_3 f_4 W(3,4|1,2)
-f_1 f_2 W(1,2|3,4)],
\nonumber
\end{flalign}
where $W(ij|kl)=(2\pi)^4\delta^4(P_{ij}-P_{kl})
\sum_{s_i, s_f}{|\mathcal{M}_{ij\to kl}|^2}$,
and $f_i$ is the phase space distribution.

Using the unitary condition $\int d\tilde{p_3}d\tilde{p_4} W(3,4|1,2)=\int d\tilde{p_3}d\tilde{p_4} W(1,2|3,4)$
to write the collision integral only in terms of the annihilation rate
\begin{flalign}
\frac{1}{1+\delta_{12}} &\int \prod^4_{i=1} d\tilde{p_i} \;
(f_3 f_4 -f_1 f_2)
W(1,2|3,4),
\end{flalign}
we keep out a statistical factor accounting for the possibility of identical particles. 

By definition, the invariant cross section, using the flux in the form~(\ref{invF1}), is
\begin{flalign}
\sigma=\frac{1}{4 (p_1\cdot p_2)v_{\text{rel}}}\int d\tilde{p_3}d\tilde{p_4} \frac{W(1,2|3,4)}{g_1 g_2},
\label{cross}
\end{flalign}
being $g_i=(2s_i+1)$ the spin degrees of freedom. 

Assuming as usual that the annihilation products are described by the equilibrium phase space distribution
at zero chemical potential $f_{0,i}$, 
we have $f_3 f_4 =f_{\text{0},3} f_{\text{0},4}=f_{\text{0},1} f_{\text{0},2}$,
the last equality following from energy conservation. Hence
\begin{flalign}
\frac{g_1 g_2}{1+\delta_{12}}\int\prod^2_{i=1} \frac{d^3 \boldsymbol{p}_i}{(2\pi)^3 E_i}(p_1\cdot p_2)(f_{\text{0},1} f_{\text{0},2} -f_1 f_2)
\sigma v_{\text{rel}}.
\nonumber
\end{flalign}
The equilibrium phase-space distribution $f_{0,i}$ is related to the
number density $n_0$ and to the momentum distribution $f_{0,p}(\boldsymbol{p})$ by
 ${g_i}/{(2\pi)^3}\,f_{0,i}=n_{0,i} f_{0,p}(\boldsymbol{p})$.
Assuming further  that the non-equilibrium phase-space function at finite chemical potential $f_i$
remains proportional to the equilibrium momentum distribution by a factor given by the non-equilibrium 
number density $n_i$, 
${g_i}/{(2\pi)^3}\, f_i =n_i f_{0,p}(\boldsymbol{p})$,
we obtain 
\begin{flalign}
\frac{1}{1+\delta_{12}}(n_{0,1} n_{0,2} -n_1 n_2)\langle \sigma v_{\text{rel}}\rangle.
\label{Cf_final}
\end{flalign}
When the species 1 and 2 are the same, it takes the usual form  $\langle \sigma v_{\text{rel}}\rangle (n^2_0 -n^2) $
with the factor 1/2 cancelled by stoichiometric coefficient appearing in the left-hand side of the complete 
kinetic equation, see for example Ref.~\cite{Cannoni:2014zqa}.

\subsection{Averaged thermal rate}

In Eq.~(\ref{Cf_final})  the general definition of relativistic thermal averaged rate is
\begin{flalign}
\langle \sigma v_{\text{rel}}\rangle=
\int\prod^2_{i=1} 
\frac{d^3 \boldsymbol{p}_i}{E_i}
(p_1\cdot p_2)
f_{0,p}(\boldsymbol{p}_1) f_{0,p}(\boldsymbol{p}_2) 
\sigma v_{\text{rel}}.
\label{svgeneral}
\end{flalign}
In the case of the relativistic Maxwell-Boltzmann-Juttner statistics, 
the momentum distribution is 
\begin{flalign}
f_{0,p} (\boldsymbol{p})=\frac{1}{4\pi m^2 T K_2(x)}e^{-\sqrt{\boldsymbol{p}^2+m^2}/T},
\end{flalign}
and as shown in Ref.~\cite{Cannoni:2013bza}, the six-dimensional integral on the right-hand side
of Eq.~(\ref{svgeneral}) reduces to
\begin{flalign}
\langle \sigma v_\text{rel}\rangle\,\,&=\int^1_0 dv_\text{rel} \mathcal{P}(v_\text{rel}) 
\sigma v_\text{rel},
\label{rel_everage_def}
\end{flalign}
where the probability distribution of $v_\text{rel}$, for example for $m_1=m_2=m$, is
\begin{flalign}
\mathcal{P}(v_{\text{rel}})=
\frac{x}{\sqrt{2}  K^2_2 (x) }
 \frac{\gamma^3_{_{\text{r}}} (\gamma^2_{_{\text{r}}} -1)}{\sqrt{\gamma_{{\text{r}}} +1}}
K_1 (\sqrt{2}x\sqrt{\gamma_{{\text{r}}} +1}).
\label{P_v_rel}
\end{flalign}
This is completely analogous to the nonrelativistic case where 
the probability distribution of $v_r$, for $m_1=m_2=m$,  is
\begin{flalign}
P(v_{{r}})=\sqrt{\frac{2}{\pi}}x^{3/2} 
{v}^2_{{r}}\, e^{-x\frac{v_{{r}}^2}{4}},
\label{Maxwell_vrel}
\end{flalign}
and the thermal average reads
\begin{flalign}
\langle \sigma_{nr} v_r\rangle_{nr}&=\int^\infty_0 dv_r P(v_r) \sigma_{nr} v_r.
\label{nr_average_def}
\end{flalign}
Given the total annihilation cross section $\sigma$ 
the product $\sigma v_{\text{rel}}$
will reduce to the nonrelativistic limit $\sigma_{nr} v_r$
and $\langle \sigma v_{\text{rel}} \rangle$  to $\langle \sigma_{nr} v_r\rangle_{nr}$
in the COF when $v_\text{rel}\sim v_r \ll 1$. 
Expressing Eq.~(\ref{rel_everage_def}) in terms of $s$ using (\ref{vrel_s}), we obtain the usual 
integral~\cite{Edsjo:1997bg,Cannoni:2013bza}  useful for  practical calculation 
\begin{flalign}
\langle \sigma_{} {v}_{\text{rel}} \rangle   &=
\frac{1}{8 T \prod_i m^2_i K_2 (x_i)}\crcr
&\times\int^{\infty}_{M^2} ds 
\frac{\lambda(s,m^2_1,m^2_2)}{\sqrt{s}}
K_1 (\frac{\sqrt{s}}{T}) \sigma,
\label{sigmav_1}
\end{flalign}
with $x_i=m_i/T$ and $M=(m_1+m_2)$.

We have recently become aware of the paper~\cite{weaver} where, 
probably for the first time, the thermal average of relativistic rates 
was discussed and it was realized that with the 
relativistic Maxwell-Boltzmann statistics formula (\ref{svgeneral}) reduces to 
a single integral over the distribution over the relative momentum.
With some algebra and change of variables it is easy to verify that for example 
Eqs.~(11b) and (12a) of~~\cite{weaver} coincide with Eqs.~(29) and (37) of 
Ref.~\cite{Cannoni:2013bza}. In Ref.~\cite{weaver} the cases
of collisions of two massive particles, two massless particles and a massive with a 
massless particles are treated separately as if different definitions of flux and cross
sections were necessary in each case. Clearly this distinction is unnecessary for 
the formulation we have given is completely general and valid in any case. We finally note
that an integral formula similar to (\ref{sigmav_1}) was also given in Ref.~\cite{Claudson:1983js}.

\subsection{No need for the M\o{}ller velocity.}

By noting that in Eq.~(\ref{invF}) the factor $(1-\boldsymbol{v}_1 \cdot \boldsymbol{v}_2)$
can cancel the same factor in the denominator of $v_\text{rel}$, the invariant 
flux can also be written in the form 
\begin{flalign}
F=n_1 n_2 \sqrt{(\boldsymbol{v}_1 - 
\boldsymbol{v}_2)^2 - 
(\boldsymbol{v}_1 \times \boldsymbol{v}_2)^2}.
\label{flux_moller}
\end{flalign}
In the textbook by Landau and Lifschits~\cite{landau} this form
is attributed to Pauli without giving any reference, while
its origin is more generally attributed to M\o{}ller~\cite{moller}.

It is interesting to look at original paper by M\o{}ller~\cite{moller}. 
With our notation, he 
wants to prove that the flux given (\ref{flux_moller}) is invariant. In order to do that he 
shows that this can be written as a product of two invariant quantities: the ratio $\frac{n_1 
n_2}{E_1 E_2}$ and the quantity $B=\sqrt{(p_1 \cdot p_2)^2 -m^2_1 m^2_2}$ and there he stops. 

The flux factor written in the form (\ref{flux_moller}) 
has  the same structure of thee nonrelativistic expression $n_1 n_2 v_r$.
Probably for this reason it has been later introduced in the literature the notion of M\o{}ller velocity
\begin{flalign}
\bar{v}&=\sqrt{(\boldsymbol{v}_1 - \boldsymbol{v}_2)^2 - 
{(\boldsymbol{v}_1 \times \boldsymbol{v}_2)^2}} 
=\frac{\sqrt{(p_1 \cdot p_2)^2 -m^2_1 m^2_2}}{E_1 E_2}.
\label{v_moller}
\end{flalign}
It is worth to stress that neither  M\o{}ller  nor Landau and
Lifschits attribute any particular meaning to Eq.~(\ref{v_moller}) and do not define it as a particular velocity,
even less as relative velocity.
Clearly $\bar{v}$ is nothing but the numerator of the formula defining $v_{\text{rel}}$
because $\bar{v}=(1-\boldsymbol{v}_1 \cdot \boldsymbol{v}_2)v_\text{rel}$, where 
the factor $(1-\boldsymbol{v}_1 \cdot \boldsymbol{v}_2)$ comes from the definition of the 
invariant flux (\ref{invF}).
Already this fact indicates that $\bar{v}$ is not a fundamental physical quantity and
overall, it is not the relative velocity, nor when the velocities are collinear.

On the contrary, in DM literature and in textbooks, when defining the flux factor for the relativistic 
invariant cross section, it is incorrectly asserted that in a frame where the velocities are collinear
the quantity $|\boldsymbol{v}_{1}-\boldsymbol{v}_{2}|$ is the relative velocity, 
while in a generic frame  is given by (\ref{v_moller}).
The form (\ref{flux_moller}) of the flux is a simple consequence of the
fundamental quantities (\ref{v_rel_Rel_def}) and (\ref{invF}), there is no new
physics or concept in it. For these reasons, and for its noninvariant and nonphysical nature,
$\bar{v}$ should not be used.

\end{document}